\documentclass[pdflatex,sn-vancouver-ay]{sn-jnl}
\usepackage{graphicx}
\usepackage{multirow}
\usepackage{xcolor}
\usepackage{textcomp}
\usepackage{amsmath}
\usepackage{longtable}
\usepackage{geometry}
\usepackage{array}

\title{\textbf{Are clinicians ethically obligated to disclose their use of medical machine learning systems to patients?}}

\author{\fnm{Joshua} \sur{Hatherley}}\email{jjh@hum.ku.dk}

\affil{\orgdiv{Center for the Philosophy of AI}, \orgname{University of Copenhagen, Denmark}}

\date{August 7th, 2024}

\abstract{It is commonly accepted that clinicians are ethically obligated to disclose their use of medical machine learning systems to patients, and that failure to do so would amount to a moral fault for which clinicians ought to be held accountable. Call this “the disclosure thesis.” Four main arguments have been, or could be, given to support the disclosure thesis in the ethics literature: the risk-based argument, the rights-based argument, the materiality argument, and the autonomy argument. In this article, I argue that each of these four arguments are unconvincing, and therefore, that the disclosure thesis ought to be rejected. I suggest that mandating disclosure may also even risk harming patients by providing stakeholders with a way to avoid accountability for harm that results from improper applications or uses of these systems.

\bigskip

\bigskip

This is a pre-print of: Hatherley, Joshua. 2024. Are clinicians ethically obligated to disclose their use of medical machine learning systems to patients? \textit{Journal of Medical Ethics}, forthcoming. \href{https://doi.org/10.1136/jme-2024-109905}{10.1136/jme-2024-109905}}

\begin{document}

\maketitle

\section{Introduction}

Machine learning (ML) systems are rapidly being developed to assist in a broad range of clinical tasks, including diagnosis, risk prediction, treatment recommendations, and patient monitoring \citep{rajpurkar2022ai, topol2019high}. The use of these systems in medicine holds significant potential to improve patient health, health equity, and medical efficiency \citep{esteva2019guide, topol2019deep}. However, the use of ML systems in medicine has also generated various ethical dilemmas and concerns \citep{arnold2021teasing,grote2020ethics,hatherley2020limits,sparrow2019promise}. Medical ML systems may exacerbate existing health inequities due to their vulnerability to algorithmic bias \citep{panch2019artificial}. Additionally, the reasoning processes of these systems are often opaque, making it difficult for users to understand how they arrive at their outputs \citep{hatherley2020limits}. These concerns have prompted discussion about clinicians' ethical obligations in communicating with patients about their use of medical ML systems \citep{sand2022responsibility, schiff2019should}, and particularly whether clinicians are ethically obligated to disclose their use of these systems to patients \citep{arnold2021teasing, astromske2021ethical, cohen2019informed}.

In the ethics and regulatory literatures, the received view appears to be that clinicians are ethically obligated to disclose their use of medical ML systems to patients, or that failure to do so would amount to a moral failing for which clinicians ought to be held accountable \citep{kiener2021artificial,lorenzini2023machine,ursin2021diagnosing,wang2024ethical}. According to Frank Ursin and co-authors, for instance, failure to disclose the use of a medical ML system “may amount to a form of deceit” \citep[2]{ursin2021diagnosing}. I will refer to this claim that clinicians are ethically obligated to disclose their use of medical ML systems as “the disclosure thesis.” 

There are four main arguments that have been, or could be used to defend the disclosure thesis. First, the “risk-based argument”, which states that clinicians are ethically obligated to disclose their use of medical ML systems because these systems generate risks to patient safety that warrant disclosure (section \ref{2}). Second, the “rights-based” argument, which states that clinicians are obligated to disclose because using a medical ML system without a patient’s knowledge violates one or more of their moral rights as patients (section \ref{3}). Third, the “materiality” argument, which states that a clinician must disclose their use of a medical ML system because this information is “material” to a patient’s decision (section \ref{4}). Fourth, the “autonomy” argument, which states that a clinician must disclose because using a medical ML system without a patient’s knowledge threatens shared decision-making and patient autonomy (section \ref{5}). 

My aim in this article is to evaluate the disclosure thesis by providing a critical assessment of each of these supporting arguments. Rather than assessing whether it would be prudent or maximally ethical for clinicians to disclose this information, however, I am interested in assessing whether clinicians that neglect to disclose their use of medical ML systems do something morally objectionable. I argue that each of these four arguments are unconvincing, and therefore, that the disclosure thesis ought to be rejected. I suggest that, in many cases, concerns used to justify the disclosure thesis pertain to instances of inappropriate design, use, or implementation of a medical ML system. Consequently, the disclosure thesis risks being used by clinicians, hospitals, AI developers, and other stakeholders as a buffer to enable them to avoid accountability for harm that results from improper applications or uses of these systems. Disclosure could also be used to socially license such inappropriate uses of these systems on the grounds that, by disclosing their every use of a medical ML system, a clinician has therefore done their due diligence with respect to the patient. Rather than protecting patients, therefore, operationalising an obligation to disclosure may actually risk harming patients.

Before defending these claims, however, it is worth noting that medical ML systems can be roughly divided into two basic types of system: “decision-support” systems and “treatment delivery” systems.  Decision-support systems are ML systems that provide information and recommendations to assist clinicians in formulating judgments and making decisions (e.g. systems that generate predictions of intensive care patients’ likelihood of 30-day mortality). In contrast, treatment delivery systems assist clinicians in directly treating patients (e.g. ML-enabled robotic surgery systems). The aim of this article is to assess whether clinicians are ethically obligated to disclose their use of medical ML systems for decision-support, which constitute the vast majority of systems that are currently available on the healthcare market \citep{lyell2021machine}.

Moreover, medical ML systems can also be divided into “locked” and “adaptive” systems. Locked systems are those whose parameters have been frozen to prevent them from learning from new data after undergoing regulatory assessment or being implemented in a clinical setting. In contrast, medical adaptive ML systems or MAMLS are those that can continue updating their parameters and improving their performance on an ongoing basis, even after being deployed in a clinical setting \citep{hatherley2023diachronic,sparrow2024should}. The aim of this article is to assess whether clinicians are ethically obligated to disclose their use of  locked medical ML systems, which are currently the only type of system that is eligible for premarket approval by prominent regulatory bodies such as the US Food and Drug Administration \citeyearpar{usfood2021artificial}.

\section{The risk-based argument}\label{2}

The first argument for the disclosure thesis is the risk-based argument. As noted above, the risk-based argument states that patients have a right to opt-in to healthcare practices that present substantial risks to their health and safety, and therefore, that clinicians are ethically obligated to disclose their use of medical ML systems because these systems generate substantial health and safety risks. According to proponents of the risk-based argument, medical ML systems present at least three specific health and safety risks that are both severe and likely enough to warrant disclosure to patients \citep{kiener2021artificial,ursin2021diagnosing}.

First, medical ML systems are vulnerable to adversarial attacks. These are cyberattacks that occur when manipulated data, known as “adversarial examples,” are intentionally inputted into an ML system to fool the system into generating incorrect classifications or predictions. According to proponents of the risk-based argument, the threat of adversarial attack presents substantial risks to patient health and safety because adversarial examples are visually indistinguishable from genuine input data, and because medical adversarial examples are especially easy to generate \citep{finlayson2019adversarial}. Moreover, they suggest that users may not be able to detect these interferences due to the opacity of many of the most popular varieties of ML systems (including deep learning systems and support vector machines). Proponents suggest, therefore, that malicious actors could easily interfere with patients’ medical treatment resulting in patient harm or death. 

Second, medical ML systems suffer from a variety of generalisation and robustness challenges \citep{futoma2020myth,freiesleben2023beyond}. For example, medical ML systems often struggle to generalise in new environments, resulting in substantial performance drops once they are implemented in a clinical setting \citep{futoma2020myth}. Moreover, medical ML systems tend to exhibit overconfidence in their predictions and classifications, and their performance can also degrade over time due to distributional shift \citep{freiesleben2023beyond,grote2023uncertainty}. Training data for medical ML systems can also lack contextual information, increasing the risk that these systems will learn dangerously misguided associations \citep{caruana2015intelligible}. These generalisation and robustness challenges can cause medical ML systems to generate incorrect or misleading outputs, which clinicians may struggle to identify due to the systems' opacity \citep{kiener2021artificial}. Clinicians may also overlook such errors due to the tendency for users to demonstrate reduced vigilance towards automated systems over time, otherwise known as automation complacency \citep{billings1976nasa}.

Finally, medical ML systems are highly susceptible to algorithmic bias. Algorithmic bias occurs “when the application of an algorithm compounds existing inequities in socioeconomic status, race, ethnic background, religion, gender, disability or sexual orientation to amplify them and adversely impact inequities in health systems” \citep[1]{panch2019artificial}. For example, many medical ML systems have been found to perform significantly better on Caucasian patients than on patients from other ethnic backgrounds \citep{adamson2018machine}. Systems designed for medical resource allocation have also been found to prioritise Caucasian patients over patients from other ethnic backgrounds \citep{obermeyer2019dissecting,samorani2022overbooked}. While this is often occurs because these systems are trained on non-representative datasets, algorithmic bias can also occur for other reasons such as confounding factors or improperly defined outcomes \citep{chen2021ethical,suresh2021framework}. Again, proponents of the risk-based argument suggest that algorithmic bias presents significant risks to patient health and safety, particularly for patients from certain ethnic backgrounds, due to the opacity of medical ML systems and the tendency for users to exhibit automation complacency \citep{kiener2021artificial}. Moreover, many algorithmic biases may only be detected at a high-level of statistical abstraction, making them even more difficult for clinicians to identify.

Consequently, proponents of the risk-based argument conclude that clinicians are ethically obligated to disclose their use of medical ML systems, along with the specific patient safety risks these systems present. However, there are three reasons that the risk-based argument is not nearly as convincing as it may first appear. 

First, some of the patient safety risks highlighted by proponents of the risk-based argument are exaggerated. For instance, the primary threat of adversarial attacks in healthcare is not that they will be used to interfere with patients’ care, but rather to commit insurance fraud \citep{finlayson2019adversarial}. While these attacks may threaten the bottom lines of hospitals and insurance agencies, therefore, they do not present any serious risks for patients. Moreover, adversarial attacks are easily detected. As Xingjun Ma and co-authors have demonstrated, even “simple detectors can achieve over 98 percent detection AUC against [these] attacks” \citep[1]{ma2021understanding}. Consequently, adversarial attacks may pose even less of a threat to patient safety than other types of cyberattacks, such as ransomware attacks, which have caused patient harm in the past by shutting down operating theatres and other hospital operations \citep{farringer2016send}. Despite this, calls for clinicians to disclose the risks of ransomware attacks to patients are notably absent. The patient safety risks currently associated with adversarial attacks, therefore, are not substantial enough to warrant disclosure to patients.

Second, rather than mandating that clinicians disclosure their use of medical ML systems with significant robustness challenges, these systems should instead play only a very narrow role in the clinical decision-making process to minimise their patient safety risks, or alternatively, they simply not be used in clinical practice at all until their clinical utility, reliability, and robustness have been clearly established. By requiring patients to opt-in to these risks, clinicians shifts the responsibility away from hospitals, regulators, and developers, who must ensure these systems are sufficiently robust for safe implementation and use in clinical practice. The problem, therefore, is not that the clinician has failed to disclose their use of a medical ML system to their patient, but rather that the system was used inappropriately, or perhaps that it was even used at all. 

While it could be argued that such precautions make the perfect the enemy of the good, various strategies that are likely to reduce patient safety risks associated with robustness limitations and algorithmic biases in medical ML systems are already available. Post-hoc explanations, such as saliency masks, can help clinicians identify whether an output has been influenced by irrelevant or misleading information. Model cards and datasheets can inform clinicians about the strengths and limitations of these systems and the training datasets used to develop them \citep{gebru2021datasheets,mitchell2019model}. Medical ML systems can undergo "silent" testing and be tailored to specific patient cohorts before clinical implementation \citep{hatherley2023diachronic,mccradden2022research,ong2023prediction}. Ongoing post-market surveillance and update protocols can detect and address distributional shifts or algorithmic biases over time \citep{feng2022clinical}. And algorithmic auditing can be conducted to detect and remove biases before implementation \citep{liu2022medical}. In short, many of these risks have the potential to be sufficiently minimised now, so long as regulators, hospitals, researchers, and clinicians are willing to devote the time and money towards these various strategies. 

Finally, the claim that clinicians are ethically obligated to disclose patient safety risks associated with algorithmic bias reflects a double standard. It is inconsistent to demand clinicians to disclose the risk of algorithmic bias without also requiring that they disclose the risks associated with their own implicit biases, which negatively impact patients based on race, gender, socioeconomic status, and other factors \citep{chapman2013physicians,fitzgerald2017implicit,hall2015implicit}.

Of course, several objections have been raised against this so-called “double standard” argument in the ethics literature. Notably, two double standard arguments are currently being debated in the medical AI ethics literature. First, a double standard with respect to the opacity of medical AI systems and the opacity of human minds \citep{gunther2022algorithmic,peters2023explainable,zerilli2019transparency}. Second, a double standard with respect to algorithmic biases and human biases \citep{kiener2021artificial,ploug2020right}. My aim in what follows is limited to addressing objections to the second of these double standard arguments. In particular, some critics argue that the double standard is justified because algorithmic bias presents greater risks to patient safety than implicit biases in human clinicians due to the opacity of medical ML systems and the tendency for users to exhibit automation complacency (discussed earlier in this section) \citep{kiener2021artificial}. Other critics argue that algorithmic biases in medical ML systems are not regulated to the same degree as implicit biases in human clinicians, which are regulated by a variety of formal and informal mechanisms including team-based decision-making practices, informed consent requirements, formal education in medical ethics and law, the risk of malpractice suits, and the potential for reputational harm \citep{ploug2020right}. However, these objections to the double standard argument are unconvincing, for two reasons.

First, it is far from clear that algorithmic biases pose greater patient health and safety risks than human clinicians’ own implicit biases, regardless of the opacity of medical ML systems or the tendency for users to exhibit automation complacency. Users exhibit a tendency to reject the outputs of algorithmic systems over their own judgments, otherwise known as “algorithmic aversion” \citep{burton2020systematic,dietvorst2015algorithm}. Indeed, algorithmic aversion is most commonly observed in users with relevant domain expertise, such as human clinicians that using medical ML systems \citep{burton2020systematic}. While opacity and automation complacency suggests that clinicians may be more likely to act on biased outputs from algorithmic systems than their own implicit biases, therefore, algorithmic aversion suggests just the opposite: that users demonstrate greater suspicion towards, along with a tendency to reject, the outputs of algorithmic systems in favour of their own judgments. It is plausible, therefore, that human clinicians are more likely to act on their own biased judgments than the biased outputs of an algorithmic system. 

Second, due to broader structural biases embedded in healthcare systems at large, it is doubtful that current regulatory mechanisms are particularly successful at reducing the negative impact of implicit biases in medicine \citep{vela2022eliminating}. Moreover, many of the formal and informal regulatory mechanisms that purport to regulate implicit biases in human clinicians apply equally to medical ML systems (or to relevant stakeholders in the ML ecosystem). Like human clinicians, for instance, the impact of medical ML systems is also regulated by team-based decision-making approaches. After all, medical ML systems for decision-support are not independent decision-makers, but sociotechnical agents that are (permanently) embedded in a collaborative decision-making process with one or more human clinicians \citep{grote2022competitors,kudina2021co}. Wherever a clinician decides to use a medical ML system while engaged in team-based decision-making themselves, the medical ML system will also become embedded in a team-based decision-making dynamic. Indeed, medical ML systems are more strongly regulated by team-based decision-making than human clinicians. For while human clinicians can make at least some medical decisions outside a team-based decision-making framework, medical ML systems for decision-support can only contribute to medical decision-making within a team-based dynamic.

Furthermore, like human clinicians, the developers of ML system are also likely to be subject to lawsuits and reputational harms where their systems exhibit algorithmic bias. For instance, Northpointe (now Equivante) received several lawsuits and sustained great reputational harm in response to allegations of racial biases in COMPAS, an ML system designed to assist in predicting parole applicants’ likelihood of recidivism \citep{angwin2022machine}. The patient safety risks of algorithmic bias, therefore, are likely to be regulated by the strong financial incentive for these organisations to be vigilant with respect to bias, since failing to do so may come at the cost of an organisation’s profits and reputation. Indeed, many chief information officers, information technology managers, and development leads already express serious concerns about losing customers, losing employees, losing revenue, increased regulatory scrutiny, lawsuits and legal fees due to algorithmic bias \citep{datarobot2022state}.

Ultimately, therefore, disclosing these risks associated with medical ML systems is unnecessary, either because these risks are not substantial enough to warrant disclosure, or because they must be addressed and minimised prior to implementing these systems in clinical practice.

\section{The rights-based argument}\label{3}

A second argument for the disclosure thesis is the rights-based argument. The rights-based argument states that clinicians are ethically obligated to disclose their use of medical ML systems because failing to do so risks violating one or more of a patient’s moral rights. In particular, the rights-based argument can be defended using the right to refuse diagnostics and treatment planning by medical ML systems (henceforth, “right to refuse”), recently advanced by Thomas Ploug and Søren Holm \citeyearpar{ploug2020right}. This version of the rights-based argument states that clinicians are ethically obligated to disclose their use of medical ML systems to patients because failing to do so may undermine their patients’ capacity to exercise their right to refuse. After all, if clinicians do not inform their patients of their use of medical ML systems, then patients have no opportunity to refuse the use of these systems. 

The right to refuse has both a strong and a weak version. The weak version states that patients have a right to refuse diagnostics and treatment planning performed entirely by a medical ML system. In contrast, the strong version states that patients have a right to refuse diagnostics or treatment planning that merely involves the use of a medical ML system. Importantly, the rights-based argument relies on the strong version of this right since human clinicians will always be involved in diagnostics and treatment planning that involves medical ML systems for decision-support.

But how is the strong right to refuse justified? Ploug and Holm argue that patients have a strong right to refuse because using a medical ML system without giving patients the opportunity to refuse violates their further right to “act on rational concerns about the future.”  More specifically, Ploug and Holm suggest that patients have rational concerns about the future if they can “provide a consistent explanation of how the society may end up in an undesirable state that corresponds with scientific evidence and the reasonable judgement of a group of informed people” \citep[111]{ploug2020right}. For example, rational concerns that a patient might have about medical AI could include “that AI systems may outmatch physicians, that AI diagnostics and treatment planning may become monopolised, and that AI systems may take control of key institutions in society” \citep[111]{ploug2020right}. Consequently, proponents of a strong right to refuse are likely to endorse the disclosure thesis. 

However, this rights-based argument is ultimately unconvincing because the right to act on rational concerns is implausibly broad and inclusive. For if we accept that patients have a right to act on rational concerns, and that this right entails a further right to refuse practices associated with those concerns, then we must also accept that patients have a right to refuse all healthcare practices about which they hold rational concerns. For example, stakeholders have long expressed a range of rational concerns about the societal impact of “managed care” in medicine (several of which also overlap with rational concerns about the societal impact of medical ML systems) \citep{feldman1998effects,kogut2020racial,mechanic1996impact,zoloth1995patient}. However, the claim that patients therefore have a right to refuse managed care practices is simply false or, at best, purely aspirational.

Moreover, the more that ML is incorporated into medical practice, the more practically infeasible the strong right to refuse would become. For example, where ML is incorporated into the operations of basic diagnostic systems (e.g. X-ray machines, MRI machines, or ECG machines), it may be overly demanding to expect hospitals to offer an alternative that does not utilise ML approaches, particularly if ML approaches are found to produce better patient health outcomes than approaches that do not utilise ML. Offering both ML and non-ML diagnostic options would require hospitals to maintain parallel systems, which is likely to be resource-intensive. The financial and logistical burden of upholding dual systems could also divert resources from other crucial areas of patient care. While Ploug and Holm suggest that, in such cases, patients right to refuse could be limited to a right to demand a reduction in the use of medical ML systems in their diagnostics and treatment planning, the underlying rationale for the strong right to refuse is nevertheless severely weakened. 

Ultimately, therefore the rights-based argument is unconvincing because a strong right to refuse diagnostics and treatment planning from medical ML systems is implausibly broad and overly demanding. 

\section{The materiality argument}\label{4}

A third argument for the claim that clinicians are ethically obligated to disclose their use of medical ML systems to patients is the materiality argument. The materiality argument states that clinicians are ethically obligated to disclose their use of medical ML systems because this information is “material” to their patients’ decisions. Roughly, information is “material” to a patient’s decision if it meets a minimum standard of relevance or significance with respect to this decision. Unlike the other three arguments discussed in this article, the materiality-based argument suggests not only that clinicians are ethically obligated to disclose their use of medical ML systems to patients, but also that failure to disclose this information violates a patient’s informed consent. 

The materiality argument has recently been defended by Jessica Findley and co-authors \citeyearpar{findley2020keeping}. Findley and co-authors claim that information is likely to be material if disclosing it causes a substantial number of patients to change their consent decisions. This is because evidence of this sort would suggest that it is unlikely that many patients would have given consent has this information been disclosed \citep{spece2014empirical}. According to Findley and co-authors, the use of a medical ML system is material to a patient’s decision because it is likely to lead a substantial number of patients to change their consent decisions. AI systems are currently subject to significant media hype and fear-mongering, and many patients are likely to have strong preferences about whether or not these systems are used during the course of their treatment. Moreover, users tend to exhibit “algorithmic aversion” towards ML systems (discussed earlier in section \ref{2}), which suggests that patients would likely avoid relying on medical ML systems over their own judgments, or the judgments of other human beings, even when these systems perform better than human decision-makers \citep{burton2020systematic,dietvorst2015algorithm}.

Consequently, proponents of the materiality argument endorse the disclosure thesis. However, there are four reasons that the materiality argument is unconvincing.

First, as I. Glenn Cohen observes, medical ML systems for decision-support play an analogous role in the clinical decision-making process as various other sources of information that are immaterial to patients’ decisions, including “vague memories from a medical school lecture, what the other doctors during residency did in such cases, the latest research in leading medical journals, the experience with and outcomes of the last 30 patients the physician saw, etc.” \citep[1442]{cohen2019informed}. If clinicians are not ethically obligated to disclose all the factors that influence their judgments or recommendations, it is unclear why they ought to disclose the influence of a medical ML system. 

Second, empirical data about patients’ preferences and behaviours cannot robustly determine what information clinicians ought to disclose to patients. Preferences are often unstable and change over time, and patients may change their consent decisions in response to irrelevant information that clinicians are not ethically obligated to disclose. Indeed, requiring that clinicians disclose certain information solely on the basis of patients’ preferences would have ethically unacceptable consequences. For example, if many patients in a homophobic society were to change their consent decisions in response to the information that their clinician was homosexual, clinicians would be ethically obligated to disclose their sexuality to patients. Empirical evidence about how people currently do behave, therefore, is insufficient to demonstrate a conclusion about how people should behave. 

Third, even if what counts as “material information” could be settled by appealing to empirical data alone, the argument fails to demonstrate that a substantial number of patients would in fact change their consent decisions in response to discovering that their clinician’s judgements have been influenced by a medical ML system. This is because, while algorithmic aversion has been observed amongst users of algorithmic systems, it has not yet been observed toward the users of these systems. In short, there is currently no evidence to suggest that people exhibit algorithmic aversion not only towards the outputs of algorithmic systems, but also toward the users of these systems themselves. 

Finally, even if a substantial number of patients do change their consent decisions in response to discovering that the outputs of a medical ML system have informed their clinician’s judgments, it is not clear that this is what a “reasonable” patient would do under similar or identical circumstances. According to the patient-based standard of materiality, material information is that which a “reasonable” patient would consider relevant or significant for their decision under similar or identical circumstances \citep{beauchamp2019principles}. However, given that algorithmic aversion is, strictly speaking, an irrational bias against algorithmic system \citep{dietvorst2015algorithm}, it seems unlikely that a “reasonable” patient would exhibit algorithmic aversion, and therefore, unlikely that they would change their consent decision after discovering that their clinician has been influenced by a medical ML system.

Ultimately, therefore, the materiality argument is also unsuccessful in demonstrating that clinicians are ethically obligated to disclose their use of medical ML systems to patients.

\section{The autonomy argument}\label{5}

The final argument for the disclosure thesis that I discuss in this article is the autonomy argument. Autonomy refers to “at minimum, self-rule that is free from controlling interference by others and from limitations, such as inadequate understanding, that prevent meaningful choice” \citep[58]{beauchamp2019principles} The autonomy argument states that clinicians are ethically obligated to disclose their use of medical ML systems because using these systems to inform one’s clinical judgments and recommendations threatens to interfere with patient autonomy. In particular, proponents of the autonomy argument suggest that medical ML systems threaten patient autonomy by virtue of their potential to compromise the ethical ideal of shared decision-making in medicine.

Shared decision-making refers to “an approach where clinicians and patients share the best available evidence when faced with the task of making decisions, and where patients are supported to consider options, to achieve informed preferences” \citep[1361]{elwyn2012shared}. Shared decision-making, as an ethical ideal, aims to protect and promote patient autonomy by striking an appropriate balance between two extremes: paternalism on one end, in which clinicians make decisions on behalf of the patient “for their own good,” and consumerism on the other, in which clinicians provide patients with all relevant factual information, but the responsibility for making the decision rests on patients themselves. 

Proponents of the autonomy argument suggest that medical ML systems threaten to compromise shared decision-making and patient autonomy in two ways. 

First, medical ML systems often contain embedded ethical values. For example, IBM’s Watson for Oncology prioritises treatments that are likely to maximise a patient’s lifespan over treatments that are most likely to maintain or improve their quality of life \citep{mcdougall2019computer}. However, some patients value the quality of their life over its duration. DreaMed Advisor Pro recommends insulin dosages that aim to achieve glycaemic control in diabetic patients \citep{usfood2018artificial}. However, pursuing glycaemic control can come at the cost of recurrent morbidity and potential mortality in these patients due to the risk of iatrogenic hypoglycaemia, which some patients may prefer to minimise or avoid \citep{cryer2014glycemic}. Finally, decision thresholds for probabilistic classifiers may also be set at levels that are misaligned with patients’ own attitudes toward the risk of false-positive or false-negative diagnoses \citep{birch2022clinical}. For instance, while some decision thresholds are likely to be optimised to minimise false positives, some patients will be more concerned about avoiding false-negatives. Since these embedded values are typically inflexible and cannot be tailored to align with the values and preferences of each individual patient, proponents of the autonomy argument suggest that these systems could interfere with patient autonomy by introducing values into the shared decision-making process that are misaligned with those of the patient \citep{mcdougall2019computer,birch2022clinical}.

Second, others suggest that the opacity of medical ML systems is likely to preclude clinicians from answering basic questions about patients’ medical treatment and care, thereby compromising the ethical ideal of shared decision-making \citep{bjerring2021artificial,holm2023justified}. For example, where patients are diagnosed with a particular condition by an opaque medical ML system, clinicians may be unable to answer certain “why-questions” about their AI-informed judgments, e.g. “Why have you diagnosed me with condition \textit{x} rather than \textit{y}?” \citep[see][]{bjerring2021artificial}. In such cases, it is suggested that clinicians’ responses would be limited to saying, “the AI said so.”

Consequently, proponents of the autonomy argument conclude in favour of the disclosure thesis for the sake of protecting patient autonomy and the integrity of the shared decision-making process. However, there are four reasons that the autonomy argument is unconvincing

First, as others have pointed out, medical ML systems containing embedded ethical values need not necessarily interfere with patient autonomy or shared decision-making so long as these systems play a merely assistive role in the clinical decision-making process \citep{duran2021afraid,di2019should}. AI-generated recommendations do not mean that clinicians no longer discuss available treatment options or their risks and benefits with patients in order to come to a shared decision. The AI system does not make decisions, but only provides information for clinicians to consider when formulating their own professional judgments and recommendations. So long as the clinician knows what values have been embedded in these systems and engages responsibly with patients in coming to a shared decision, the threat to patient autonomy and shared decision-making from embedded values in these systems is negligible. 

Second, embedded ethical values are contained in various other sources of information that play an analogous role in clinical decision-making to medical ML systems. As noted earlier, medical ML systems play a role in decision-making that is analogous to “vague memories from a medical school lecture, what the other doctors during residency did in such cases, the latest research in leading medical journals, the experience with and outcomes of the last 30 patients the physician saw, etc.” \citep[1442]{cohen2019informed}. A medical textbook that recommends prioritising glycaemic control over minimising the risk of iatrogenic hypoglycaemia contains embedded ethical values in the same way as DreaMed Advisor Pro. Moreover, clinicians themselves have their own decision thresholds at which they diagnose patients with certain illnesses, and these thresholds differ from clinician to clinician. However, this does not obligate clinicians to disclose the influence of specific medical textbooks to patients, nor indeed any other instance in which their judgments or recommendations are influenced by sources of evidence that contain embedded ethical values. Nor is it clear that any of these factors pose any greater risk to patient autonomy than embedded ethical values in medical ML systems.

Third, while I think it is doubtful that opacity in medical ML systems undermines clinicians’ capacity to the extent that they will be unable to answer basic why-questions about their patients’ diagnoses and treatment, disclosing the use of a medical ML system would do little to address these concerns in the first place were they, in fact, sound. For even if the clinician discloses their use of an opaque medical ML system, they would still be unable to explain the reasoning behind their judgments or recommendations. They would still be limited to saying, “the AI told me so.” Again, the problem here does not appear to be that the clinician did not disclose their use of a medical ML system to their patient, but rather that the output of this system played an outsized role in the clinical decision-making process, or that the system was implemented in clinical practice in the first place. 

Ultimately, therefore, disclosing the use of medical ML systems to address these concerns with patient autonomy is misguided and unnecessary. As with the risk-based argument, this is either because these risks are not substantial enough to warrant disclosure, or because they must be addressed and minimised prior to implementing these systems in clinical practice.

\section{Conclusion}\label{6}

According to the disclosure thesis, clinicians are ethically obligated to disclose their use of medical ML systems. Writers in the ethics literature tend to endorse the disclosure thesis on the basis of four distinct arguments: the risk-based argument, the rights-based argument, the materiality argument, and the autonomy argument. In this article, have argued that each of these arguments is unconvincing for a variety of reasons, and therefore, that the disclosure thesis ought to be rejected. 

In many cases, moreover, the arguments used to defend the disclosure thesis often appeal to concerns that are ultimately about (the potential for) improper use, premature implementation, or flawed design of these systems. One implication of this is that the disclosure thesis risks being used by clinicians, hospitals, AI developers, and other stakeholders as a buffer to enable them to avoid accountability for harm that results from improper applications or uses of these systems. It could also be used to socially license such inappropriate uses of these systems on the grounds that, by disclosing their every use of a medical ML system, a clinician has therefore done their due diligence with respect to the patient. Rather than protecting patients, therefore, operationalising an obligation to disclosure may actually risk harming patients. 

\section*{Acknowledgments}

Thanks to Rob Sparrow, Justin Oakley, Paul Formosa, Sven Nyholm, Rosa Martorana, Courtney McMahon, Tommy Ness, Lauritz Aastrup Munch, the audience at Aarhus University’s Spring Ethics Workshop, and two anonymous referees for their helpful comments on earlier drafts of this article. The work for this article was supported by a Research Training Program Scholarship from the Australian Government and a Carlsberg Foundation Young Researcher Fellowship (CF20-0257).

\bibliography{main}
\end{document}